\documentclass[twocolumn,superscriptaddress,aps,pra,nofootinbib,floatfix]{revtex4-2}
\usepackage{amssymb}
\usepackage{amsmath}
\usepackage{graphicx}
\usepackage{dcolumn}
\usepackage{bm}
\usepackage{dsfont}
\usepackage[breaklinks=true,colorlinks,citecolor=blue,linkcolor=blue,urlcolor=blue]{hyperref}
\usepackage[caption=false]{subfig}
\usepackage{physics}
\usepackage{mleftright}
\usepackage{bbm}
\usepackage{soul}
\DeclareMathOperator{\sinc}{sinc}
\usepackage{tikz,xcolor}
\definecolor{lime}{HTML}{A6CE39}
\DeclareRobustCommand{\orcidicon}{%
	\begin{tikzpicture}
	\draw[lime, fill=lime] (0,0) 
	circle [radius=0.16] 
	node[white] {{\fontfamily{qag}\selectfont \tiny ID}};
	\draw[white, fill=white] (-0.0625,0.095) 
	circle [radius=0.007];
	\end{tikzpicture}
	\hspace{-2mm}
}
\foreach \x in {A, ..., Z}{%
	\expandafter\xdef\csname orcid\x\endcsname{\noexpand\href{https://orcid.org/\csname orcidauthor\x\endcsname}{\noexpand\orcidicon}}
}

\begin{document}
\title{Phonon-photon conversion as mechanism for cooling and coherence transfer}
\date{\today}
\author{Alessandro Ferreri\orcidA{}}
\email{a.ferreri@fz-juelich.de}
\affiliation{Institute for Quantum Computing Analytics (PGI-12), Forschungszentrum J\"ulich, 52425 J\"ulich, Germany}
\affiliation{Theoretical Quantum Physics Laboratory, RIKEN, Wako-shi, Saitama 351-0198, Japan}
\author{David Edward Bruschi\orcidB{}}
\email{david.edward.bruschi@posteo.net}
\affiliation{Institute for Quantum Computing Analytics (PGI-12), Forschungszentrum J\"ulich, 52425 J\"ulich, Germany}
\affiliation{Theoretical Physics, Universit\"at des Saarlandes, 66123 Saarbr\"ucken, Germany}
\author{Frank K. Wilhelm\orcidC{}}
\affiliation{Institute for Quantum Computing Analytics (PGI-12), Forschungszentrum J\"ulich, 52425 J\"ulich, Germany}
\affiliation{Theoretical Physics, Universit\"at des Saarlandes, 66123 Saarbr\"ucken, Germany}
\author{Franco Nori\orcidD{}}
\affiliation{Theoretical Quantum Physics Laboratory, RIKEN, Wako-shi, Saitama 351-0198, Japan}
\affiliation{Center for Quantum Computing, RIKEN, Wako-shi, Saitama 351-0198, Japan}
\affiliation{Physics Department, The University of Michigan, Ann Arbor, Michigan 48109-1040, USA}
\author{Vincenzo Macrì\orcidE{}}
\email{macrivince1978@gmail.com}
\affiliation{Theoretical Quantum Physics Laboratory, RIKEN, Wako-shi, Saitama 351-0198, Japan}
\affiliation{Dipartimento di Ingegneria, Universit\`{a} degli Studi di Palermo, Viale delle Scienze, 90128 Palermo, Italy}

\begin{abstract}
The dynamical Casimir effect is the physical phenomenon where the mechanical energy of a movable wall of a cavity confining a quantum field can be converted into quanta of the field itself. This effect has been recognized as one of the most astonishing predictions of quantum field theory. At the quantum scale, the energy conversion can also occur incoherently, namely without an physical motion of the wall. By means of quantum thermodynamics, we show that this phenomenon can be employed as a tool to cool down the wall when there is a non-vanishing temperature gradient between the wall and the cavity. At the same time, the mechanism responsible for the heat-transfer enables to share the coherence from one cavity mode, driven by a laser, to the wall, thereby forcing its coherent oscillation. Finally, we show how to employ one laser drive to cool the entire system including the case when it is composed of other subsystems.
\end{abstract}
\maketitle

\section{Introduction}
Quantum thermodynamics is the modern branch of physics that extends and adapts the principles and  main concepts of thermodynamics to the quantum scale \cite{ gemmer2009quantum,Vinjanampathy2016, strasberg_first_2021, kurizki_kofman_2022}. In contrast to its classical counterpart, one of the revolutionary accomplishments in quantum thermodynamics is the realization of thermal machines based on single quantum systems instead of large ensembles of particles \cite{quan_quantum_2005, quan_quantum_2007,quan_quantum_2009,PhysRevResearch.2.032062,abah_single-ion_2012,zhang_dynamical_2022}. In general, it is expected that the employment of such systems will lead to both the miniaturization of the working substance, and the optimization of the work extraction \cite{scully_extracting_2003, korzekwa_extraction_2016,perarnau-llobet_extractable_2015,PhysRevE.100.062140, PhysRevE.102.062123, campisi_power_2016}. 

All quantum systems interact with their own environment, and exchange energy with it. The mathematical formalism that best tackles problems in this context, and therefore allows to quantify the heat flows between the system and the heat-baths, is the theory of open quantum systems \cite{kosloff_quantum_1984, breuer2002theory,kosloff_quantum_2013}. Among the successful applications in this direction, can be found studies of quantum heat engines and quantum refrigerators based on continuous devices via master equations \cite{kosloff_quantum_1984, levy_quantum_2012, kosloff_quantum_2014}. In fact, proven the thermodynamic equivalence between continuous and discrete-strokes devices \cite{uzdin_equivalence_2015}, the master equation approach is largely used because it offers a detailed description of the dynamics in reaching the steady state, and its regime of validity is extended also to out-of-equilibrium scenarios. 

The presence of quantum features in mesoscopic objects, along with the possibility to control the interaction with the environment, makes cavity optomechanical systems an interesting platform for the study of quantum thermodynamics \cite{mari_quantum_2015, PhysRevA.100.022501, fong_phonon_2019}, as well as valid candidates for the realization of quantum heat engines 
\cite{zhang_quantum_2014, zhang_theory_2014, brunelli_out--equilibrium_2015, dong_work_2015, kolar_extracting_2016,PhysRevLett.124.210601, PhysRevResearch.5.043274}.
Cavity optomechanics studies the quantum interactions occurring in cavity systems between the confined electromagnetic field and the cavity walls (also called mechanical resonators) by means of radiation pressure \cite{law_interaction_1995,Nunnenkamp2011,meystre_short_2013,Aspelmeyer2014}. The wall can in principle oscillate and, when the amplitude of oscillations is sufficiently small, the displacement from the mean position can be represented by a bosonic degree of freedom \cite{Aspelmeyer2014}.
Mechanical oscillators interacting with cavity mode in the quantum regime are systems with great potential in many field of research, as quantum metrology and sensing \cite{aspelmeyer2010quantum}, offering a route towards fundamental tests of quantum mechanics in a hitherto inaccessible parameter regime of size and mass \cite{romero2011large}. Furthermore, they have been widely used to study quantum protocols for creation and control of mechanical quantum superpositions of macroscopic objects \cite{Bose1999,Marshall2003,Pepper2012,Garziano2015,Macri2016}. 


In general, the standard assumption in optomechanics is that the electromagnetic field can be described by a single optical mode, which is highly populated by photons prepared in a coherent state (i.e., the quantum state of a laser beam). This approach permits to focus the attention on the fluctuation of the number of excitations only, thereby linearizing the Hamiltonian and drastically reducing the complexity of the interaction \cite{Aspelmeyer2014}. However, in the last decade the interest in optomechanical Hamiltonians beyond the linear regime has increased \cite{he_quantum_2012}, and nonlinear effects such as the radiation pressure on the movable wall \cite{qvarfort_enhanced_2019,qvarfort_master-equation_2021} and the so called ``Casimir terms" 
\cite{law_interaction_1995, butera_field_2013, macri_nonperturbative_2018, PhysRevA.106.033502, Russo2023} are often taken into account. The name of the latter stems from the interpretation of the dynamical Casimir effect in optomechanical frameworks \cite{dalvit_fluctuations_2011, dodonov_fifty_2020}, in which the squeezing of the quantum vacuum derives from the energy conversion between the mechanical and the optical quantum degrees of freedom.

This work explores the quantum thermodynamics of cavity optomechanical systems in which the Casimir terms, i.e., the phonon-photon conversion, can be seen as enabling a quantum channel for the propagation of heat between subsystems. The setup under consideration consists of a cavity characterized by a movable mirror, whose position fluctuates, and it is described as a bosonic degree of freedom, namely as a quantum harmonic oscillator. The cavity confines a quantum field, and we assume that only two modes participate to the dynamics. The wall and the cavity as a whole are coupled to two different baths, each at different temperatures: the cavity is coupled to a cold bath at temperature $T_{\textrm{c}}\simeq 0$, whereas the wall exchanges heat with a hot bath at temperature $T_{\textrm{w}}>0$. Finally, we assume that the first cavity mode is driven by an external laser characterized by the same frequency of the cavity mode. 

The frequency of the mirror is tuned to be twice the frequency of the lower cavity mode, thereby activating a resonance that enables the flow of excitations between the two subsystems. The presence of a second cavity mode allows us to investigate the effects of the interactions on the dynamics of higher-order resonant modes. We see that the flows of particle between the wall and the cavity mode induce the cooling of the cavity wall as well as the motion of the wall, the latter a consequence of the up-conversion of coherent photons into coherent phonons.

The paper is structured as follows: in Sec.~\ref{model} we introduce the system and the Hamiltonian, emphasizing the role of every interaction terms. In Sec.~\ref{master equation} we present both the formalism for the study of the dynamics, namely the master equation in the dressed picture, and the quantities of interest for the comprehension of both the dynamics and the thermodynamics. In Sec.~\ref{numres} we expose and discuss our numerical results. We conclude in Sec.~\ref{conclu}. Some details are left for the appendix. More precise, in Appendix~\ref{appA} we analyze the coherence transfer processes, showing analytically the origin of the oscillation frequency of both the wall and the second mode.
\section{Theoretical model}\label{model}
\begin{figure}[ht!]
	\centering
\includegraphics[width=0.9\linewidth]{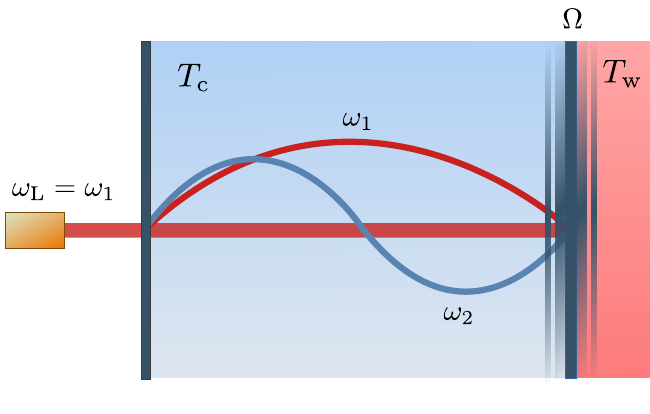}
	\caption{Pictorial representation of the system: a cavity with a free-moving mirror confines a quantum scalar field (the first two modes are here depicted). The field is in contact with a cold bath at temperature $T_\textrm{c}\simeq0$, whereas the movable wall of the cavity is couple to a bath at $T_\textrm{w}>0$. A laser at the same frequency of mode 1 coherently excites the first cavity mode.}
	\label{scheme}
\end{figure}

The system consists of a cavity confining a (1+1)dimensional uncharged massless scalar field.  This approximation well describes the Transverse Electric (TE) modes of the electromagnetic field confined in a three-dimensional box when the parameters of the system are opportunely tuned \cite{crocce_quantum_2002,PhysRevResearch.5.043274}. Importantly, we assume that the cavity possesses a movable wall that interacts with the field via a position-dependent interaction. The system is depicted in Fig.~\ref{scheme}.

The setup considered here provides a well-known mathematical model for optomechanical systems, wherein the motion of the light-massive vibrating wall is associated to the zero-point fluctuation of a quantum harmonic oscillator \cite{Aspelmeyer2014}. The presence of a movable wall therefore leads to the existence of an additional quantum degree of freedom representing the small quantized vibration of the wall.
The Hamiltonian of such system can be derived from first principles following different procedures \cite{law_interaction_1995, PhysRevA.106.033502}, and to each harmonic oscillator one associates a fundamental frequency $\omega$, and annihilation operators $\hat{a}$ and $\hat{a}^\dag$ that satisfy the canonical commutation relations $[\hat{a},\hat{a}^\dag]=1$. Operators of different modes commute. It is well-known that the spectrum of the field is discrete, where each level is labelled by the  quantum number $n$, which is a positive natural number. Exciting mode $n$ to its $m$-the energy level requires $m\hbar \omega_n$ quanta of energy. Therefore, any action performed on the cavity has a probability, albeit perhaps small, to excite any mode. Nevertheless, for the sake of simplicity and for the scope of this work, we assume that we will effectively be able to truncate the spectrum of our system to the first two modes only, thereby ignoring the remaining modes of the quantum scalar field. 

Combining all together, it can be shown that the Hamiltonian of the system reads $\hat H_\textrm{s}=\hat H_0+\hat H_\textrm{I}$, where $\hat H_0=\hbar\omega_1\hat a_1^\dag\hat a_1+\hbar\omega_2\hat a_2^\dag\hat a_2+\hbar\Omega\hat b^\dag\hat b$ is the noninteracting (free) Hamiltonian and
\begin{align}
\hat H_\textrm{I}=&\hbar\epsilon\frac{\sqrt{\omega_1\omega_2}}{2}\left(\hat a_1+\hat a_1^\dag\right)\left(\hat a_2+\hat a_2^\dag\right)\left(\hat b+\hat b^\dag\right)\nonumber\\
&+\hbar\epsilon\sum_{j=1,2}\frac{\omega_j}{2}\left(\hat a_j+\hat a_j^\dag\right)^2\left(\hat b+\hat b^\dag\right)
\label{Hamiltinian:interaction}
\end{align}
is the Hamiltonian of interaction. In the expressions above, we have that $\omega_j$, $\hat a_j$ and $\hat a_j^\dag$ are the frequency and the operators of the cavity mode labelled by $j=1,2$, whereas $\Omega$, $\hat b$ and $\hat b^\dag$ are the frequency and the operators of the movable wall. From here on, Latin subscripts such as $i,j$ will always label the cavity modes and therefore take values $1,2$. Finally, $\epsilon:=dL/L$ is the dimensionless oscillation amplitude of the wall. For later convenience, we also define the dimensional coupling parameters, $g_{11}=\epsilon \, \omega_1/2$, $g_{22}=\epsilon \, \omega_2/2$ and $g_{12}=\epsilon \, \sqrt{\omega_1\omega_2}/2$.

The Hamiltonian in Eq.~\eqref{Hamiltinian:interaction} expresses the interactions between the movable wall and the cavity modes that arise in optomechanics. Among all possible contributions we: (i) recognize the radiation pressure terms $\hat a_j^\dag\hat a_j(\hat b+\hat b^\dag)$, which are largely studied in quantum optomechanics, namely in regime wherein $\Omega\ll\omega_j$ and the cavity contains a large number of photons \cite{aspelmeyer2012quantum, Aspelmeyer2014}. Note that, three types of phonon-photon conversion terms are present and can become relevant in appropriate regimes: (ii) the single mode photon up- and down-conversion terms $\hat a_j^{\dag2} b+\hat a_j^2\hat b^\dag$; (iii) the two-mode photon up- and down-conversion term $\hat a_1^\dag\hat a_2^\dag\hat b+\hat a_1\hat a_2\hat b^\dag$; (iv) the Raman scattering term $\hat a_i\hat a_j^\dag (\hat b+\hat b^\dag)$. Any of these terms can become dominant whenever specific frequency resonances between the cavity mode and the movable wall are activated \cite{butera_field_2013,roelli2016molecular,macri_nonperturbative_2018,Butera2019,Neuman2019,Russo2023}. For instance, the up- and down-conversion mechanism occurs if $\Omega=\omega_i+\omega_j$, namely when the frequency of the wall is equal to either twice $\omega_j$ (single-mode) or the sum of the two cavity frequencies (two-mode), whereas the Raman scattering process occurs if $\Omega=\omega_2-\omega_1$.
The interaction Hamiltonian in Eq.~\eqref{Hamiltinian:interaction} also contains: (v) the counter-rotating terms $\hat a_i\hat a_j \hat b+\hat a_i^\dag\hat a_j^\dag \hat b^\dag$. Beyond their contribution to the energy shift related to fluctuations of the quantum vacuum \cite{butera_field_2013}, their presence becomes crucial for the observation of higher-order processes \cite{Russo2023,Montalbano2023}.

\section{The master equation in the dressed picture}\label{master equation}

The three subsystems, i.e., the two cavity modes and the wall, interact in a strong regime. Therefore, it is convenient to study its dynamics by means of the master equation in the dressed picture \cite{beaudoin_dissipation_2011,settineri2018}. Before doing this, we introduce the transition amplitudes for the canonical position operators evaluated on the dressed basis: $u^{(n)}_{ij}=\langle i\lvert (\hat a_n+\hat a_n^\dag)\rvert j\rangle$ and $w_{ij}=\langle i\lvert (\hat b+\hat b^\dag)\rvert j\rangle$, where $\rvert i\rangle$ is the $i$-th eigenstate of the Hamiltonian $\hat H_\textrm{s}$ with eigenenergy $E_i$, and $n=1,2$.
To calculate the quantities of interest, we establish a set of dressed annihilation operators for all subsystems, namely $\hat A_n=\sum_{j,i>j}u^{(n)}_{ij}\hat P_{ij},\,\,\hat B=\sum_{j,i>j}w_{ij}\hat P_{ij}$,
as well as the transition operators $\hat P_{ij}=\lvert i\rangle\langle j\rvert$. Note that the operators $\hat P_{ij}$ are not projectors since $\hat P_{ij}^2=0$, but the diagonal operators $\hat P_{jj}=P_{ij}^\dag\hat P_{ij}=P_{ji}\hat P_{ji}^\dag=\lvert j\rangle\langle j\rvert$ are.

We add an additional term $\hat H_\textrm{d}(t)$ in the Hamiltonian that has the effect of coherently exciting the lowest cavity mode $j=1$. This external drive term is time dependent and its oscillation frequency is chosen to match the frequency $\omega_1$ of this mode. It represents the presence of a laser entering into the cavity and transferring coherence to the corresponding mode. This additional term can be written in terms of the dressed operators, and it reads
\begin{align}
\hat H_\textrm{d}(t)=F\left(e^{-i\tilde\omega_{1}t}\hat A_1+e^{i\tilde\omega_{1}t}\hat A_1^\dag\right),
\end{align}
where $F$ describes the intensity of the laser. As soon as the laser intensity is much less than the damping rates of both cavity and wall, we can claim that the laser does not alter the eigenstates of the Hamiltonian $\hat H_\textrm{s}$, therefore it can be treated perturbatively. We can now write the total Hamiltonian of the system as
\begin{align}
\hat H_{\textrm{tot}}(t)=&\hat H_\textrm{s}'+\hat H_\textrm{d}(t)
\end{align}
with the system Hamiltonian expressed with respect to its own eigenbasis $\hat H_{\textrm{s}}'=\sum_iE_i\hat P_{ii}$. 
To compact the nomenclature, henceforth we will simply write $\hat H_{\textrm{tot}}$, having clear that it is time-dependent.

Usually, the three subsystems would be coupled to different baths. 
In several works it was demonstrated that quantum systems can interact with a non-Markovian (possibly colored \cite{JingPRA13,CornQIP16}) common bath displaying revival of entanglement \cite{PlenioPRA99,LiPRB08,Minganti2021}, which can be enhanced by the presence of measurement protocols \cite{ManiscalcoPRL2008,LiJP09}, and revealing nontrivial dynamics \cite{FrancicaPRA09,Macr2022}. From now on, we will always consider the two modes of the cavity sharing a common bath at $T_{\textrm{c}}\simeq 0$ (unless explicitly stated in particular cases), with damping rate $\kappa$, while the movable mirror will be coupled to a different bath with damping rate $\gamma$ and temperature $T_{\textrm{w}}>0$.
Under these assumptions, the master equation in the dressed picture reads
\begin{equation}
\frac{d\hat{\rho}}{dt}=-i[\hat H_{\textrm{tot}}',\hat{\rho}]+\hat{\mathcal{L}}_c(\hat{\rho})+\hat{\mathcal{L}}_w(\hat{\rho}),
\label{me:appendix}
\end{equation}
where we have defined the differential generators
\begin{align}
    \hat{\mathcal{L}}_c(\hat{\rho})=&\frac{\kappa}{4}\sum_{j,i>j}\lvert u^{(1)}_{ij}+u^{(2)}_{ij}\rvert^2\,\hat{D}_{ij}[\hat{\rho}],\nonumber\\
    \hat{\mathcal{L}}_w(\hat{\rho})=&\frac{\gamma}{2} \sum_{j,i>j}\lvert w_{ij}\rvert^2\,\hat{D}_{ij}[\hat{\rho}],
\end{align}
with the meanig of $\hat{D}_{ij}[\hat{\rho}]=n_{ij}(T)\mathcal{D}[\hat P_{ij}]\hat{\rho}+(1+n_{ij}(T))\mathcal{D}[\hat P_{ji}]\hat{\rho}$.
These terms contain the thermal excitation numbers $n_{ij}(T)=(e^{\hbar (\omega_i -\omega_j)/k_{B}T}-1)^{-1}$, and the superoperators 
\begin{align}
\mathcal{D}[\hat P_{ij}]\hat{\rho}=\frac{1}{2}\bigl(2\hat P_{ij}\hat{\rho}\hat P_{ij}^\dag-\hat{\rho}\hat P_{ij}^\dag\hat P_{ij}-\hat P_{ij}^\dag\hat P_{ij}\hat{\rho}\bigr).
\label{DP}
\end{align}

By employing these operators and numerically solving the dressed master equation in the Schrödinger picture, we can explore the driven-dissipative system dynamics. 
In particular, we are interested in the time evolution of both the population and the quadrature position of all subsystems. The former is achieved by evaluating the average value of the particle number for both the optical and the mechanical modes, $N_n(t)=\textrm{Tr}[\hat A_{n}^\dag\hat A_{n}\hat\rho(t)]$ and $N_{{\text{w}}}(t)=\textrm{Tr}[\hat B^\dag\hat B\hat\rho(t)]$, respectively. On the other hand, the quadrature position operators allow to describe: (i) the actual motion of the wall via $X_{{\text{w}}}(t)=\textrm{Tr}[(\hat B^\dag+\hat B)\rho(t)]$, (ii) the amplitude of the field-modes $X_j(t)=\textrm{Tr}[(\hat A_{j}^\dag+\hat A_{j})\rho(t)]$.

In order to investigate the quantum thermodynamic features of the system, we introduce

\begin{align}
\hat{\mathcal{L}}_{\text{c}}^*(\hat H_{\textrm{tot}}')=&\frac{\kappa}{4}\sum_{j,i>j}\left\lvert u^{(1)}_{ij}+u^{(2)}_{ij}\right\rvert^2\,\hat{D}_{ij}[\hat H_{\textrm{tot}}'],\nonumber\\
  \hat{\mathcal{L}}_{\text{w}}^*(\hat H_{\textrm{tot}}')=&\frac{\gamma}{2} \sum_{j,i>j}\lvert w_{ij}\rvert^2\,\hat{D}_{ij}[\hat H_{\textrm{tot}}']
\end{align}
with the following simplified notation  $\hat{D}_{ij}[\hat H_{\textrm{tot}}']=n_{ij}(T)\mathcal{D}[\hat P_{ij}]\hat H_{\textrm{tot}}'+(1+n_{ij}(T))\mathcal{D}[\hat P_{ji}]\hat H_{\textrm{tot}}'$ for this superoperator. 

Finally, we are able to determine the heat-flow from the cold bath to the cavity, composed by the two optical modes, by means of the expression $\mathcal{J}_{\text{c}}(t)=\textrm{Tr}[\hat{\mathcal{L}}_{\text{c}}^*(\hat H_{\textrm{tot}}')\rho(t)]$, and the heat-flows from the hot bath to the wall by means of the expression $\mathcal{J}_{\text{w}}(t)=\textrm{Tr}[\hat{\mathcal{L}}_{\text{w}}^*(\hat H_{\textrm{tot}}')\rho(t)]$. The power produced is given by the expression $\mathcal{P}(t)=\textrm{Tr}[\hat{\dot{H_\textrm{d}}}(t)\rho(t)]$.
\section{Numerical results}\label{numres}

We investigate here the thermodynamic features of the system. To obtain our goal, we look at the heat flows $\mathcal{J}_{\text{c}}(t)$ and $\mathcal{J}_{\text{w}}(t)$, as well as at the laser power $\mathcal{P}(t)$. We stress that, while the cavity consists of two optical modes, the heat propagation mediated by $\mathcal{J}_{\text{c}}(t)$ involves the whole cavity.
Note that, the positive sign of both the heat-rates and the laser power indicates the energy absorption from the source: a positive heat-rate (laser power) means that the relative subsystem is absorbing heat (coherence), or thermal (coherent) excitations from its own bath (the laser), whereas a negative sign is a signature of the release of energy. Concerning the dynamics, we focus our attention on both the population of the cavity modes $N_j(t)$ and the movable mirror $N_{{\text{w}}}(t)$, as well as their coherent modulation. The latter is calculated by averaging the position quadratures $X_{j}(t)$ and $X_{{\text{w}}}(t)$.

A main role in our discussion is played by the resonance condition $\Omega=2\omega_1$. We will always assume that this resonance is active throughout the dynamics. The main goal of this work is in fact to show that this resonance, which leads to the phonon-photon conversion mechanism, allows for the existence of a valid channel for both the heat and the coherence transfer through the various subsystems. Another important element of our analysis is the strong coupling regime, which is determined by the coupling constants $g_{ij}$. These parameters are tuned in such a way that they are large enough to guarantee an efficient interaction between the subsystems, while simultaneously being smaller than any mode frequencies \cite{frisk_kockum_ultrastrong_2019}. For this reason, the hierarchy between the various parameters will be $\omega_{j},\Omega\gg g_{jj}>\gamma>\kappa\gg F$, with exception of Sec.~\ref{diffloss}.

Finally, we stress that henceforth we will distinguish the bare from the dressed frequency of the first cavity mode, $\tilde\omega_1$. The correction follows the fact that the resonant interaction between the subsystems leads to both shift and splits of the eigenenergies, slightly altering also the effective bare frequencies of the Hamiltonian. By numerically calculating the eigenenergies in proximity of the resonant frequency $\omega_1=\Omega/2$, we correct it to the value that minimizes the energy split, namely $\tilde\omega_1$. Including this small energy shift means optimizing the resonant exchange of excitation between the subsystem \cite{PhysRevResearch.5.043274}.

\subsection{Dynamics in presence of thermal gradient and thermal equilibrium}

\subsubsection{Heat flows} 
\begin{figure}[ht!]
	\centering
 \includegraphics[width=1\linewidth]{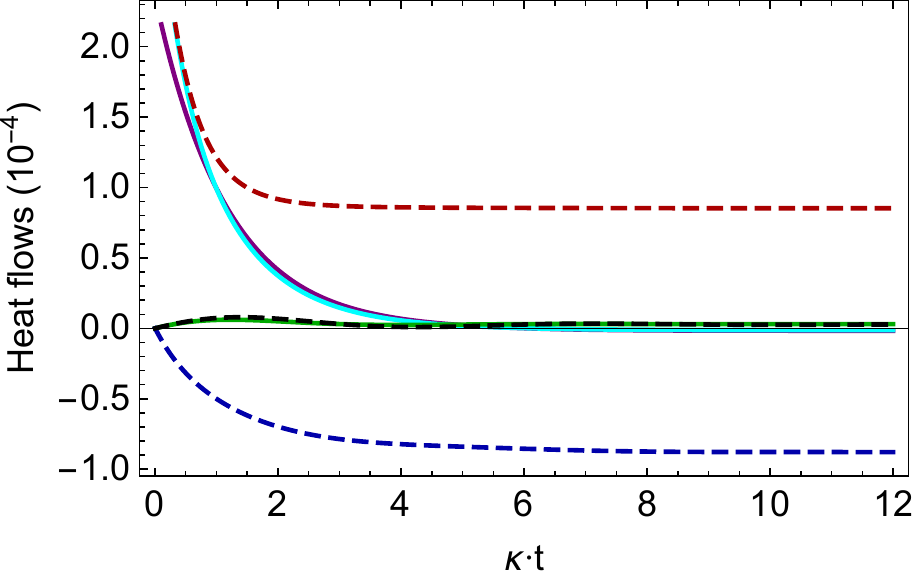}
	\caption{Time evolution of the heat flows and the laser power. By setting $T_{\textrm{c}}=T_{\textrm{h}}=0.3$, we plot the heat flows of the mirror (cyan solid line), of the cavity (purple solid line), and of the laser power (green solid line). By setting $T_{\textrm{c}}=10^{-6}$ and $T_{\textrm{w}}=0.3$, we also plot the heat flows of the mirror (red dashed line), of the cavity (blue dashed line),  and of the laser power (black dashed line). Other parameters are: $\tilde\omega_{1}=0.502$, $\Omega=\omega_{2}=1$, $\epsilon=0.05$, $\gamma=0.009$, $\kappa=0.003$, $F=0.02\gamma$. Frequencies and temperatures are normalized with respect to $\omega_{2}$.}
	\label{grvseqh}
\end{figure}

In order to demonstrate that the phonon-photon conversion is a valid mechanism for the heat transfer between the mechanical and the optical modes, we first look at the heat-flows between the cavity, and the wall, including also the contribution of the laser, in two different scenarios. In a first case, we assume that the temperature of the two baths is the same, $T_{\textrm{c}}=T_{{\text{w}}}>0$. Note that the two cavity modes share a common bath. In the second case, we set a thermal gradient between the two baths, namely, $T_{{\text{w}}}>0$ and $T_{{\text{c}}}\simeq0$. The results are plotted in Fig.~\ref{grvseqh}. 

\begin{figure}[ht!]
	\centering
	\includegraphics[width=1\linewidth]{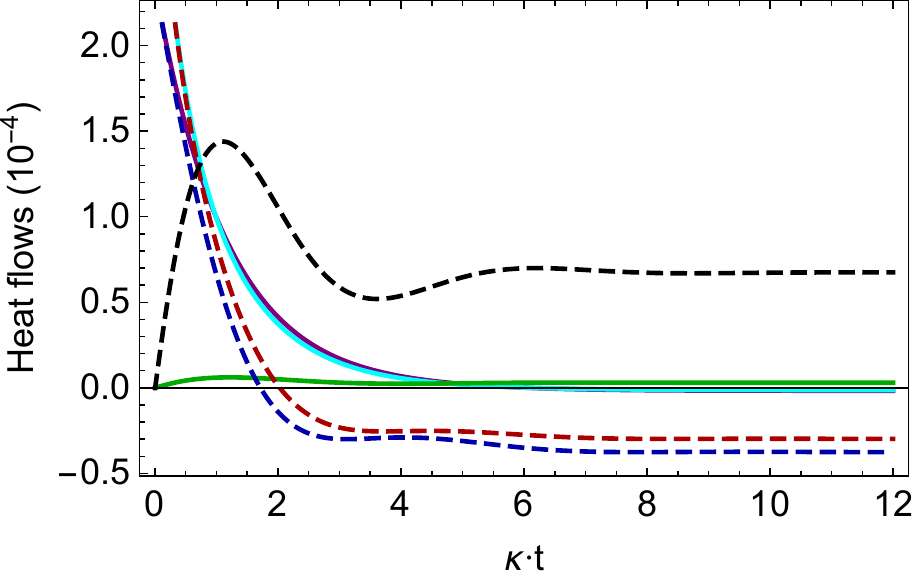}
	\caption{Time evolution of the heat flows and the laser power in thermal equilibrium, $T_{\textrm{c}}=T_{\textrm{h}}=0.3$. By setting $F=0.02\gamma$, we plot the heat flows of the mirror (cyan solid line), of the cavity (purple solid line), and of the laser power (green solid line). For $F=0.1\gamma$, we also report the heat flows of the mirror (red dashed line), of the cavity (blue dashed line),  and of the laser power (black dashed line). Other parameters are: $\tilde\omega_{1}=0.502$, $\Omega=\omega_{2}=1$, $\epsilon=0.05$, $T_{\textrm{c}}=T_{\textrm{h}}=0.3$, $\gamma=0.009$, $\kappa=0.003$. Frequencies and temperatures are normalized with respect to $\omega_{2}$.}
	\label{StrongLaser}
\end{figure}

In the first scenario, we let the system be initialized in its ground state, observing heat absorption from the environment during the dynamics. However, since there is no temperature gradient, the heat flows drastically lessen during the evolution until they reach zero, indicating that the entire system is thermalized with the two baths. Once the system approaches the steady state we observe that both heat flows become negative. This is due to the presence of the laser, which is able to cool down the two subsystems while performing some work on the first mode, despite its modest intensity. To amplify this effect, we performed a further simulation with a higher laser intensity, as shown in Fig.~\ref{StrongLaser}. We see that, while absorbing power from the laser, the system releases heat to the environments, therefore cooling all subsystems down at the same time. The fact that both the cavity and the wall release heat to their own baths at the same time is the first evidence that the resonance between wall and cavity mode establishes a valid channel for the heat flow. 

Now we simulate the case in which a temperature gradient between the two baths is different from zero. In this case, the heat flows behave in a very different manner. Indeed, when this condition applies the wall absorbs phonons from its own bath. At the same time, however, due to the resonant coupling with the cavity, part of these ``\emph{hot}" phonons are converted into photons, and are then finally released to the cavity bath at $T_{{\text{c}}}\simeq0$. The graph in Fig.~\ref{grvseqh} shows that, in the presence of the temperature gradient, the heat flow of the wall is always positive, indicating a continuous absorption of heat from its own bath. At the same time, the heat flow of the cavity is negative, meaning that it releases heat to its own \emph{cold} bath. Since the only source of thermal excitations is the bath coupled to the wall, this dynamics describes a heat-flow between the two baths promoted by the resonant interaction between the mechanical and the first optical modes. Physically, this permanent heat-flow maintains the entire system out of thermal equilibrium.

\subsubsection{Populations}
\begin{figure}[ht!]
	\centering
	\includegraphics[width=1\linewidth]{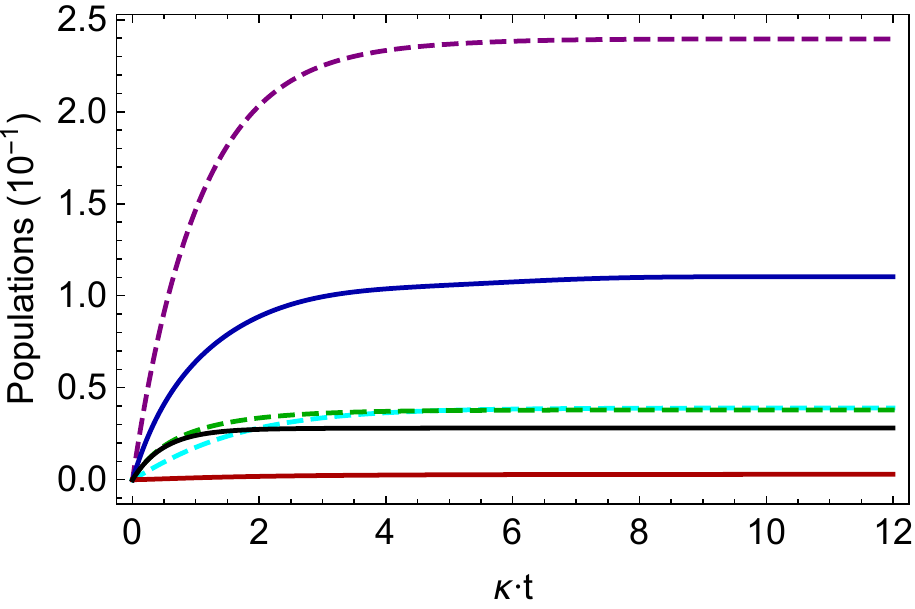}
	\caption{Time evolution of the populations. By setting $T_{\textrm{c}}=T_{\textrm{h}}=0.3$, we plot the populations of the mirror (green dashed line), of the cavity mode 1 (purple dashed line), and of the cavity mode 2 (cyan dashed line). By setting $T_{\textrm{c}}=10^{-6}$ and $T_{\textrm{w}}=0.3$, we also plot the populations of the mirror (black), of the cavity mode 1 (blue), and of cavity mode 2 (red). Other parameters are: $\tilde\omega_{1}=0.502$, $\Omega=\omega_{2}=1$, $\epsilon=0.05$, $\gamma=0.009$, $\kappa=0.003$, $F=0.02\gamma$. Frequencies and temperatures are normalized with respect to $\omega_{2}$.}
	\label{grvseqp}
\end{figure}

We now compare the time-evolution of the vibrational and optical populations for the two scenarios, namely, thermal equilibrium and thermal gradient. Our results are plotted in Fig.~\ref{grvseqp}. The graph shows that all modes get highly populated as soon as the two subsystems, namely the cavity and the wall, are coupled to two different baths at the same temperature. Since we fixed $\omega_2=\Omega$, it does not come as a surprise that the second cavity mode and the mechanical mode are very similarly populated once the system reaches the steady state. Indeed, the population of each subsystems at the end of the dynamics is close to what is expected by the Bose-Einstein statistics. For instance, $N_{{\text{w}}}(t_\textrm{f})=0.038$ and $N_{{\text{w}}}^\textrm{(BE)}=(e^{\hbar \omega/(k_{B} T)}-1)^{-1}=0.037$, with $t_\textrm{f}$ is the time during which the system has reached the steady-state. The small discrepancy stems from the fact that we are dealing with an interacting resonant system, therefore the eigenenergies of the Hamiltonian do not correspond to the bare energies of the single quantum harmonic oscillators due to the presence of  energy dressing \cite{butera_field_2013,PhysRevResearch.5.043274}.

Once we impose the gradient between cavity and wall, namely by tuning the temperature of the bath coupled to the cavity at $T_{{\text{c}}}\simeq 0$, the resonant particle exchange between wall and cavity automatically sets the system out of equilibrium. As we can see from Fig.~\ref{grvseqp}, in this scenario the phononic degree of freedom is still populated thanks to the absorption of particles from its own bath at finite temperature. However, a significant part of these excitations are converted into photon pairs populating the resonant cavity mode, whereas a smaller fraction of phonons are responsible for the excitation of the second cavity mode due to higher order resonances \cite{PhysRevA.100.022501}. Indeed, assuming that the second mode and the wall have the same frequency, the excitation of the second mode is related to a second-order effective Hamiltonian of the form $\hat H_{\textrm{I}}^{\textrm{eff}}\propto \epsilon^2(\hat a_1^\dag\hat a_1(\hat a_2^\dag)^2\hat b^2+\textrm{h.c.})$,  responsible for the induced phonon-photon conversion.

We see that the resonant interactions between the wall and the two cavity modes affect the population of the phononic mode at the steady state, which turns out to be smaller than what we observed in thermal equilibrium scenario. Indeed, according to our numerical analysis, the phononic population in the out-of-equilibrium scenario amounts to $N_{{\text{w}}}(t_\textrm{f})=0.028$, namely the 26\% less than what is observed in thermal equilibrium. To this population we can approximately associate an effective temperature $T_{{\text{w}}}^{\textrm{eff}}$ given by inverting the Bose-Einstein statistics, thus obtaining $T_{{\text{w}}}^{\textrm{eff}}\simeq 0.277$, which is slightly less than the temperature of the bath. Although this effective temperature does not vary significantly from the temperature of the bath, $T_{{\text{w}}}=0.3$, in the next section we will show that it is possible to further cool down the wall by appropriately manipulating specific cavity parameters.

\begin{figure}[ht!]
	\centering
	\includegraphics[width=1\linewidth]{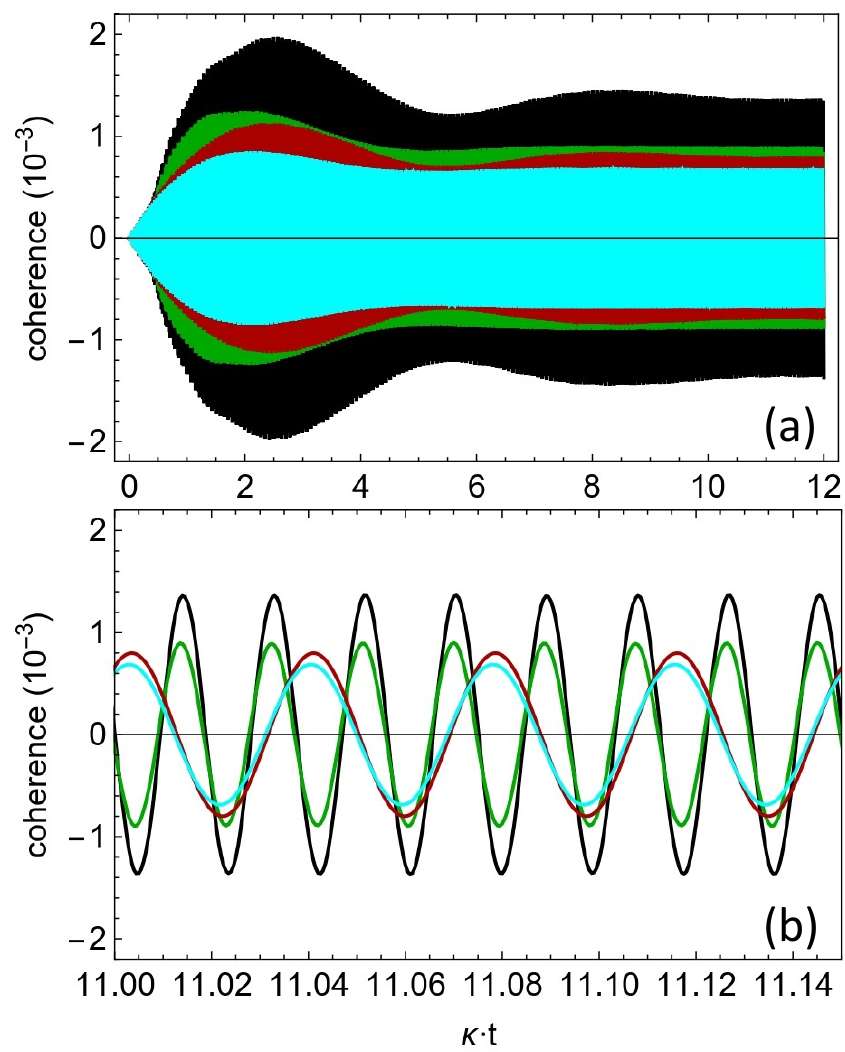}
	\caption{(a) Time evolution of the position quadrature operator of the mirror and mode 2. (b) An enlarged view of the latter on the range $\kappa\cdot t\in\{11,11.14\}$, is provided. By setting $T_{\textrm{c}}=T_{\textrm{w}}=0.3$, one sees coherence of the mirror (green line), and cavity mode 2 (cyan line), while, by setting $T_{\textrm{c}}=10^{-6}$ and $T_{\textrm{w}}=0.3$ one sees coherence of the mirror (black line), the  and cavity mode 2 (red line). Other parameters are $\tilde\omega_{1}=0.502$, $\Omega=\omega_{2}=1$, $T_{\textrm{c}}=10^{-6}$, $T_{\textrm{w}}=0.3$, $\epsilon=0.05$, $\gamma=0.009$, $F=0.02\gamma$. Frequencies and temperatures are normalized with respect to $\omega_{2}$.}
 \label{grvseqcz}
\end{figure}


\subsubsection{Coherences}
So far we have discussed the purely thermodynamic aspects of the system. However, the presence of the laser driving the first mode offers further perspectives which can be taken into account. In fact, we now want to explore the possibility of employing the same quantum channel, namely the channel enabled by the resonant terms of the Hamiltonian, thereby inducing the wall to move coherently. 

Results comparing the dynamics of the quadrature position operators for the wall and the second mode in both thermal equilibrium and the out-of-equilibrium scenarios are plotted in Fig.~\ref{grvseqcz}. Since the first cavity mode is directly excited by the laser, the oscillation amplitude of $X_{1}(t)$ is always much higher than the oscillation amplitudes of both $X_{\textrm{w}}(t)$ and $X_{2}(t)$, and is therefore of no interest.

The graph shows that the average values of the quadrature operators are not zero, meaning that the second optical mode and the mechanical mode are not only excited by the presence of the hot bath, but also coherently. Clearly, these oscillations must stem as a consequence of the input energy from the laser that drives the first cavity mode, since this is the only source of coherence in the system.

The fact that the wall oscillates does not come as a surprise: the wall is directly resonant with the pumped mode, therefore the coherence transfer between the two modes must be related to the up-conversion of coherent photons into phonons. The interesting fact is that it oscillates with its own frequency, namely twice the frequency of the laser. This is due to the fact that the frequency of the laser and frequency of the pumped mode coincide, thereby activating the resonant phonon-photon conversion between the laser and the wall. A off-resonant laser would induce the wall to oscillate at twice the frequency of the laser, namely $2\omega_\textrm{L}$, but with an expected much lower amplitude. More details about these aspects are discussed in the Appendix~\ref{appA}.

On the other hand, a fundamental clue for the explanation of the coherence in the second cavity mode is provided by zooming in the oscillation of the two modes. Indeed, we observe that they oscillate at different wavelengths, although they are set at the same frequency. From Fig.~\ref{grvseqcz}\textcolor{blue}{b}, it is evident that the optical mode oscillates twice slower than the mechanical mode, namely, at the same frequency of the laser. In order to investigate this aspect, we studied the dynamics of the quadrature operators analytically, assuming for simplicity an unitary evolution (see Appendix~\ref{appA}). This analysis shows that the presence of coherence on the second mode is a consequence of the structure of the Hamiltonian. In particular, it is due to a mixture between the resonant photon-phonon conversion term (cavity mode 1 and the wall) and the three-boson coupling, namely the first line of Eq.~\eqref{Hamiltinian:interaction}. Interestingly, we observe that the oscillation frequency of the second mode does not depend on its own frequency, suggesting that the presence of the wall induces all cavity modes to oscillate at the same frequency of the driven mode (with lower amplitude for higher mode numbers). 

Once the temperature gradient is introduced, we observe an increase of the oscillation amplitude in both the wall and second cavity mode 2. This amplification of the oscillations reveals that the system partially cleans up the motion from the thermal fluctuations, and slightly improves its coherence.

\begin{figure}[ht!]
	\centering
	\includegraphics[width=1\linewidth]{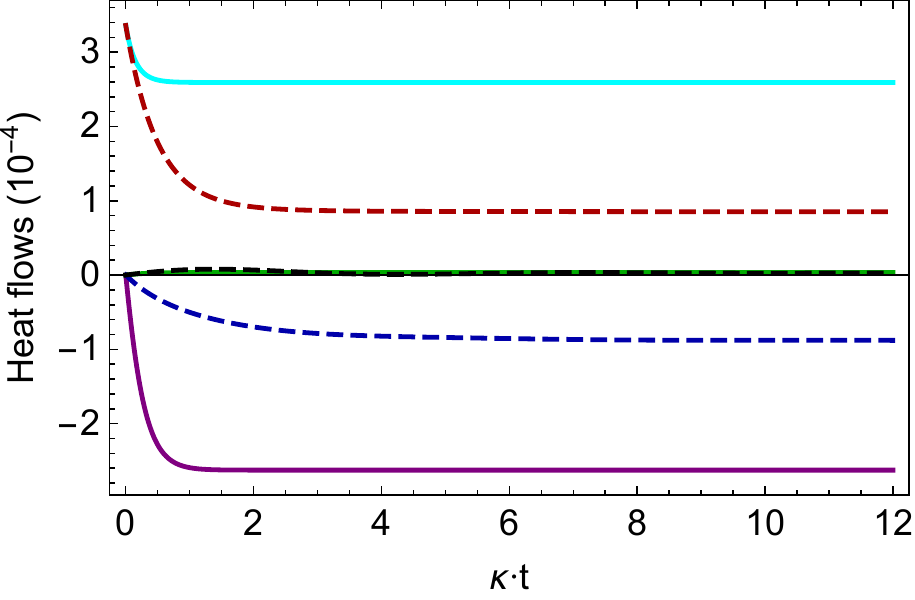}
	\caption{Time evolution of the heat flows and the laser power, using $\kappa=0.003$ as time scale. By setting $\kappa=0.03$, we plot heat flows of the mirror (cyan solid line), of the cavity (purple solid line), and of the laser power (green solid line). By setting $\kappa=0.003$, we also plot heat flows of the mirror (red dashed line), of the cavity (blue dashed line), and of the laser power (black dashed line). Other parameters are: $\tilde\omega_{1}=0.502$, $\Omega=\omega_{2}=1$, $T_{\textrm{c}}=10^{-6}$, $T_{\textrm{w}}=0.3$, $\epsilon=0.05$, $\gamma=0.009$, $F=0.02\gamma$. Frequencies and temperatures are normalized with respect to $\omega_{2}$.}
	\label{hlvsllh}
\end{figure}

\begin{figure}[ht!]
	\centering
	\includegraphics[width=1\linewidth]{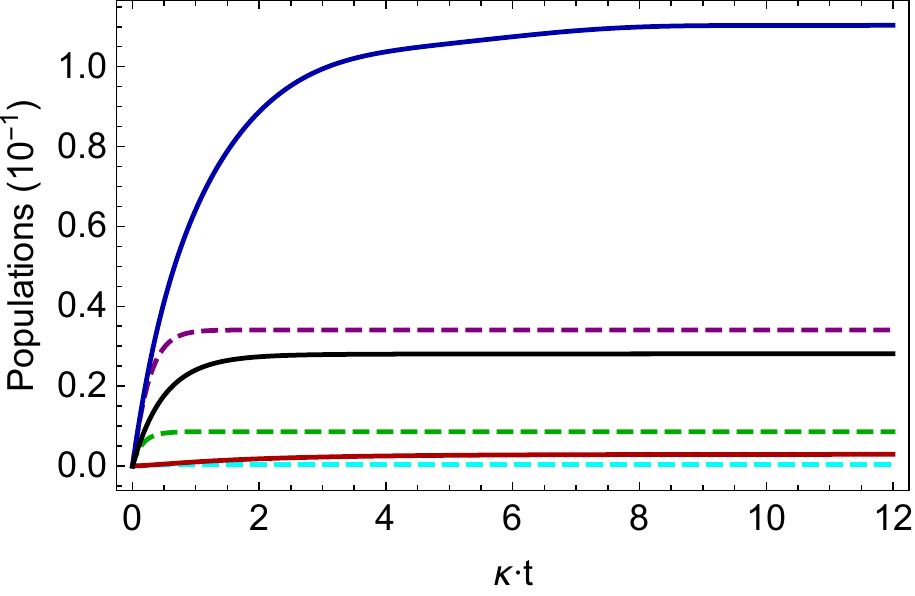}
	\caption{Time evolution of the populations, using $\kappa=0.003$ as time scale. By setting $\kappa=0.03$, we plot the populations of the mirror (green dashed line), of the cavity mode 1 (purple dashed line), and of the cavity mode 2 (cyan dashed line). By setting $\kappa=0.003$, we also plot populations of the mirror (black solid line), of the cavity mode 1 (blue solid line), and of the cavity mode 2 (red solid line). Other parameters are: $\tilde\omega_{1}=0.502$, $\Omega=\omega_{2}=1$, $\gamma=0.009$, $\epsilon=0.05$, $T_{\textrm{c}}=10^{-6}$, $T_{\textrm{w}}=0.3$, $F=0.02\gamma$. Frequencies and temperatures are normalized with respect to $\omega_{2}$.}
	\label{hlvsllp}
\end{figure}

\subsection{Different cavity losses}\label{diffloss}
We have demonstrated that the phonon-photon conversion mechanism can work as a channel that enable the heat-flows between cavities and wall. Now, we want to check whether it is possible to further cool down the wall by controlling the damping rate of the cavity, rather than directly tuning the parameters of the wall. We recall that the dynamics are occurring in the strong coupling regime, therefore internal interactions between the subsystems are more favorable than the exchanges with the baths. By improving both the damping rate of the cavity and taking advantage of the strong coupling, we can expect that the wall effectively interacts more with the cavity bath than with its own bath. 
To test this claim, we now work in the regime with a different hierarchy of the parameters: $g_{jj}>\kappa>\gamma\gg F$, namely a strong coupling regime with higher cavity losses. 

\subsubsection{Heat flows, populations and coherences}
We start by looking at the heat-flows as done before. The time evolution plotted in Fig.~\ref{hlvsllh} evidently shows a drastic enhancement of the heat flows between the two subsystems. Indeed, due to the higher damping, the wall releases excitations to the cavity bath more efficiently, forcing the wall to absorb heat from its own bath faster.

\begin{figure}[ht!]
	\centering
	\includegraphics[width=1\linewidth]{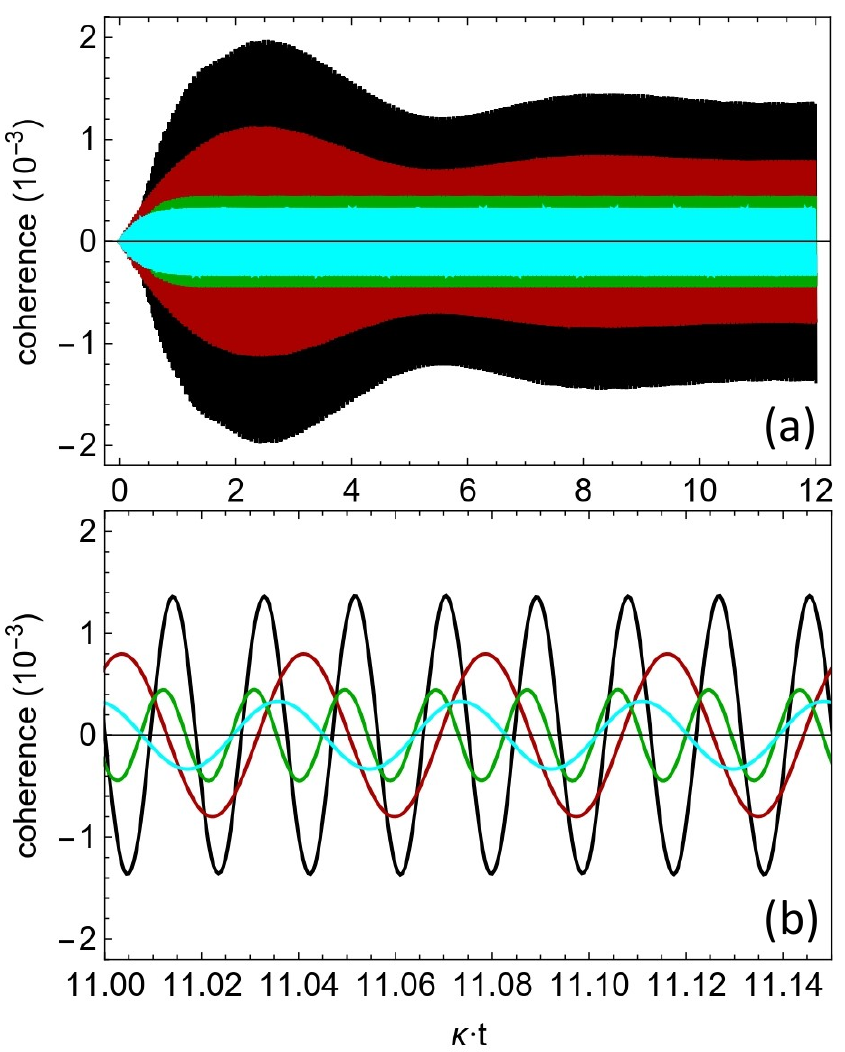}
	\caption{(a) Time evolution of the quadrature position operator of the mirror and mode 2 using $\kappa=0.003$ as time scale. (b) A large view of of the latter on the range $\kappa\cdot t\in\{11,11.14\}$. By setting $\kappa=0.03$, we show coherence of the mirror (green line) and of the cavity mode 2 (cyan line). By setting $\kappa=0.003$, we also show coherence of the mirror (black line) and of the cavity mode 2 (red line). Other parameters are: $\tilde\omega_{1}=0.502$, $\Omega=\omega_{2}=1$, $T_{\textrm{c}}=10^{-6}$, $T_{\textrm{w}}=0.3$, $\epsilon=0.05$, $\gamma=0.009$, $F=0.02\gamma$. Frequencies and temperatures are normalized with respect to $\omega_{2}$.}
 \label{hlvsllcz}
\end{figure}


Having increased only the damping rate of the cavity, which is coupled to a cold bath, and not the damping rate of the wall, we expect that the system will generally contain less excitations. This is expected because they are released to the cold bath more easily. This is exactly what we observe in Fig.~\ref{hlvsllp}, where the population of both cavity modes and the wall are plotted. Interestingly, this graph shows a net reduction of the phonon number, now amounting to $N_{{\text{w}}}(t_\textrm{f})=0.0086$, namely, about the 77\% less than what expected in thermal equilibrium. Furthermore, assuming again that the steady state of the wall approximately corresponds to the state of a quantum harmonic oscillator prepared in thermal equilibrium, we can attribute an effective temperature $T_{{\text{w}}}^{\textrm{eff}}$ by inverting the Bose-Einstein statistics, and obtain $T_{{\text{w}}}^{\textrm{eff}}\simeq 0.21$. This value is 30\% less than the temperature of the bath, confirming that the wall is cooled down. Despite the improvements in cooling down the wall, a modification of the cavity damping rate negatively affects the coherence for the total system, as can be seen in Fig.~\ref{hlvsllcz}. Indeed, the cavity mode tends to interact more with its own bath losing coherence.
\begin{figure}[ht!]
	\centering
	\includegraphics[width=1\linewidth]{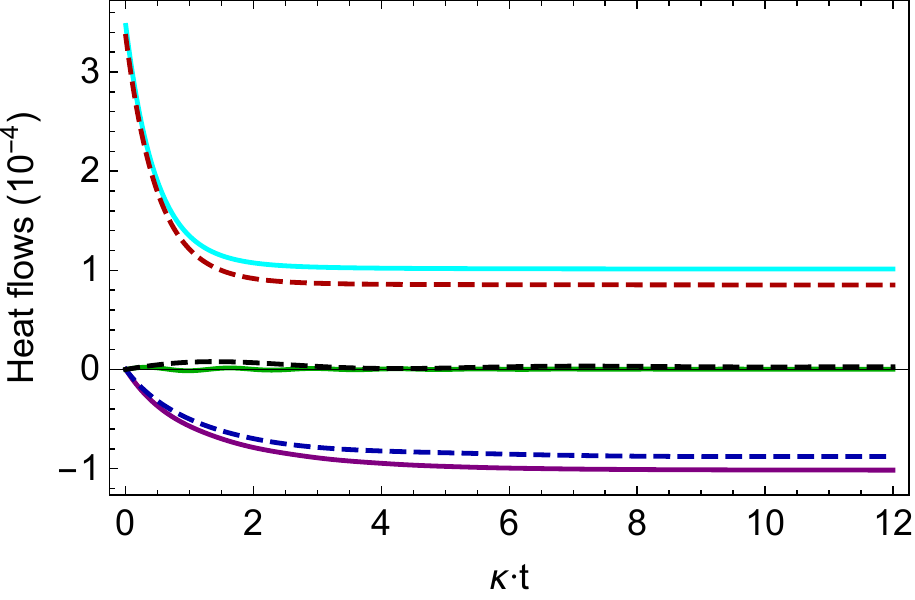}
	\caption{Time evolution of the heat flows and the laser power. By setting $\epsilon=0.05$, we show heat flows of the mirror (red dashed line) , of the cavity (blue dashed line), and of the laser power (black dotted line). By setting $\epsilon=0.1$, we show also heat flows of the mirror (cyan solid line), of the cavity (purple solid line),  and of the laser power (green solid line). Other parameters are: $\tilde\omega_{1}=0.502$, $\Omega=\omega_{2}=1$, $T_{\textrm{c}}=10^{-6}$, $T_{\textrm{w}}=0.3$, $\gamma=0.009$, $\kappa=0.003$, $F=0.02\gamma$. Frequencies and temperatures are normalized with respect to $\omega_{2}$.}
	\label{lcvshch}
\end{figure}

\begin{figure}[ht!]
	\centering
\captionsetup{format=plain}
	\includegraphics[width=1\linewidth]{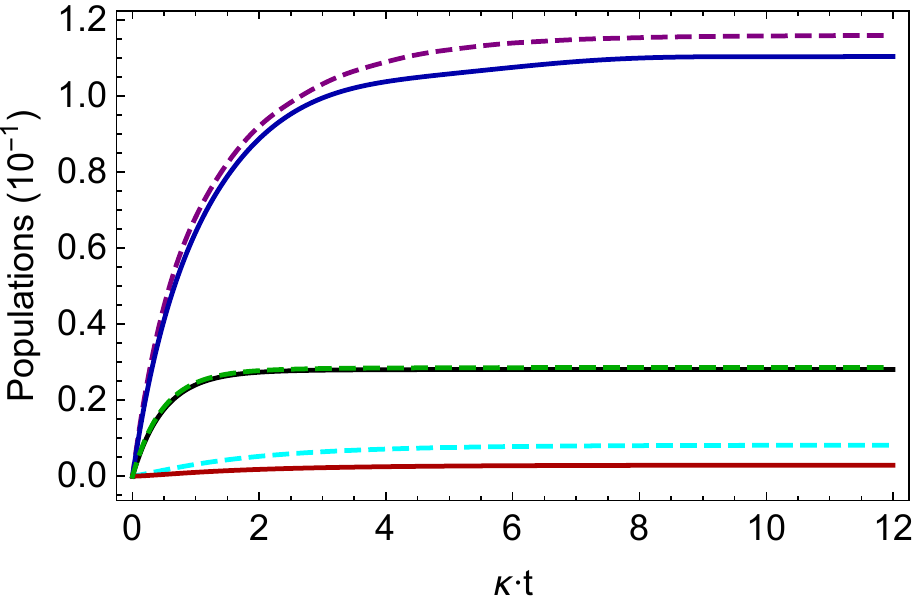}
	\caption{Time evolution of the populations. By setting $\epsilon=0.1$, we show populations of the mirror (green dashed line), of the cavity mode 1 (purple dashed line), and of the cavity mode 2 (cyan dashed line). By setting $\epsilon=0.05$, we also show populations of the mirror (black solid line), of the cavity mode 1 (blue solid line), and of the cavity mode 2 (red solid line). Other parameters are: $\tilde\omega_{1}=0.502$, $\Omega=\omega_{2}=1$, $T_{\textrm{c}}=10^{-6}$, $T_{\textrm{w}}=0.3$, $\gamma=0.009$, $\kappa=0.003$, $F=0.02\gamma$. Frequencies and temperatures are normalized with respect to $\omega_{2}$.}
	\label{lcvshcp}
\end{figure}
\begin{figure}[ht!]
	\centering
	\includegraphics[width=1\linewidth]{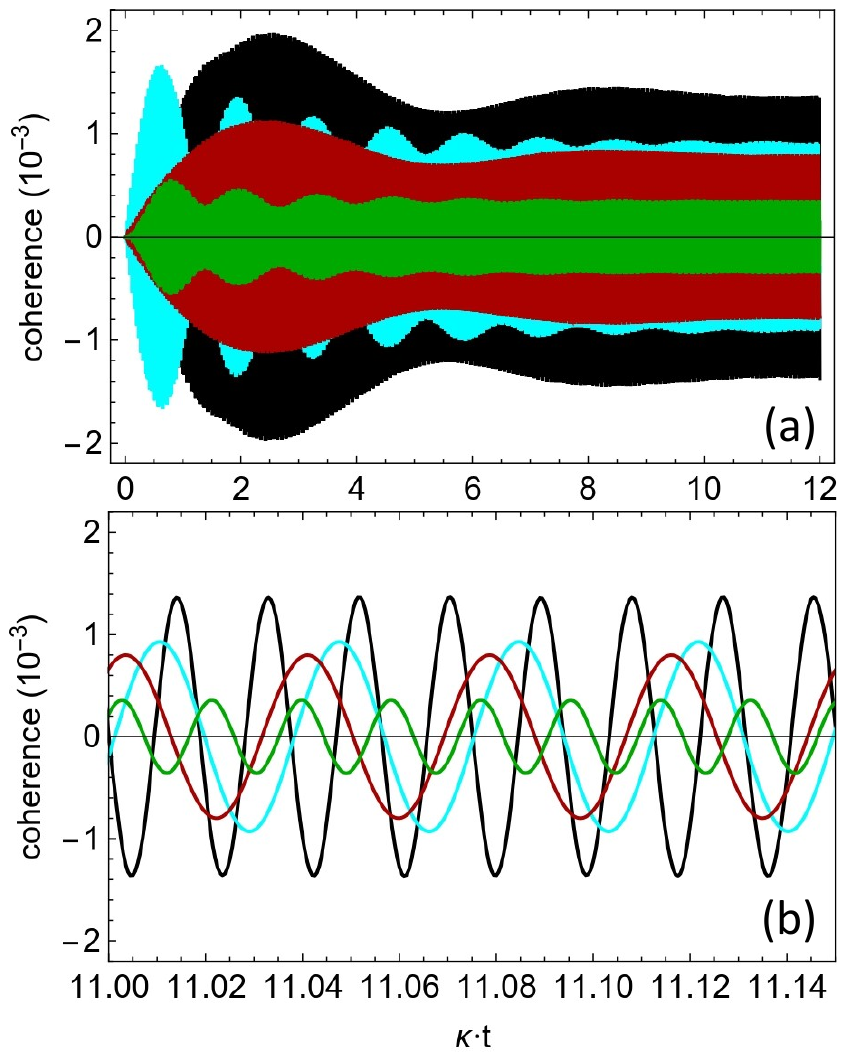}
	\caption{(a) Time evolution of the quadrature position operator of the mirror and mode 2. (b) In a large view on the range $\kappa\cdot t\in\{11,11.14\}$, by setting $\epsilon=0.1$, we show coherence of the mirror (green line), and of the cavity mode 2 (cyan line). By setting $\epsilon=0.05$, we also show coherence of the mirror (black line),  and of the cavity mode 2 (red line). Other parameters are: $\tilde\omega_{1}=0.502$, $\Omega=\omega_{2}=1$, $T_{\textrm{c}}=10^{-6}$, $T_{\textrm{w}}=0.3$, $\gamma=0.009$, $\kappa=0.003$, $F=0.02\gamma$. Frequencies and temperatures are normalized with respect to $\omega_{2}$.}
 \label{lcvshccz}
\end{figure}


\subsection{Different couplings}
As a last scenario, we want to analyze the driven-dissipative system dynamics by tuning the coupling constant $\epsilon$. We first look at the heat-flows showing the results in  Fig.~\ref{lcvshch}. The graphs show that the enhancement of the internal interactions would lead the wall to cool down. To balance this effect, the wall absorbs more heat from the hot bath. This means that the hot bath provides excitations to the wall, these excitations are converted faster into photons, and finally they are released to the cold bath with the usual rate. 

Since the damping rates do not change, for higher couplings we should expect a general increase of the population within the cavity. Indeed, in Fig.~\ref{lcvshcp} we observe a general enhancement of the optical excitations, whereas the phonon population does not undergo any relevant variation. This is a consequence of the strong coupling regime: the wall absorbs phonons from its own bath until it reaches the phonon number in accord with the temperature of the bath. At the same time, the system up- and down-converts these thermal excitations more efficiently. Therefore, we achieve a higher population of the modes with lower temperature at the price of the same damping rates. 

Although we generally observe an enhancement of the optical populations, the higher coupling and therefore the enhancement of the conversion rate  stimulate the first cavity to effectively interact more with the baths and less with the laser. This leads the mode $\omega_1$ to absorb less coherence from the laser, and consequently, less coherence is also observed in the wall. However, as one can see in Fig.~\ref{lcvshccz}, the coherence in the second mode does not change remarkably but only slightly increases. This is due to the higher variability of $X_{\textrm{2}}(t)$ as a function of the parameter $\epsilon$:  in Appendix~\ref{appA} we employ a unitary evolution and therefore do not include the coupling with the baths, nevertheless, we can analytically estimate that $X_{\textrm{2}}(t)$ depends on $\epsilon$ not linearly (see Eq.~\eqref{osc2new}), in contrast to what it is expected for $X_{\textrm{w}}(t)$ (see Eq.~\ref{oscwal}). It is therefore reasonable to think that $X_{\textrm{w}}(t)$ is in fact more sensitive to the enhancement of the coupling constant, and that it can increase its amplitude by slightly bypassing the loss effects due to the interactions with the cold bath.

\section{Conclusions}\label{conclu}
In this paper, we have explored the phonon-photon conversion mechanism as a possible quantum channel for both the propagation of heat and coherence between the movable wall and the confined electromagnetic quantum field. We have shown that the presence of the Casimir terms in the Hamiltonian allows for heat flows from a hot bath (coupled to the wall) to a cold bath (coupled to the cavity), thereby cooling down the wall. To amplify the cooling effect, we manipulated the cavity losses, and therefore strengthened the interaction with the cold bath as well as supported the heat flow.

Interestingly, we observe that the wall starts to oscillate at its own frequency, namely, twice the frequency of the laser/driven first cavity mode, whereas the second cavity mode follows the first one (i.e., same oscillating frequency), with an amplitude that is comparable to the oscillation amplitude of the wall. 
We showed that the time evolution of the second cavity mode does not oscillate with its own frequency, suggesting that all modes of the cavity fields oscillate at the frequency of the driven mode. Beyond the coherence transfer, we found that a stronger drive of one cavity mode can also cool down the entire system by releasing heat to both environments.

This work brings to light, as one result, the fact that the Hamiltonian description of optomechanical systems beyond the linearization can have multiple uses in quantum thermodynamics. 
A natural extension to this work includes the possibility to explore phenomena emerging from additional resources, such as further potential resonances due to the presence of more interacting optical and mechanical modes. This can help model more refined experiments that will be able to probe higher-order interaction terms, and therfore richer dynamics.

To conclude, we believe that our results can be of support for the realization of future quantum thermal machines based on cavity-optomechanics.

\acknowledgments
A.F. thanks the research center RIKEN for the hospitality. A.F. acknowledges the ``JSPS Summer Program 2022'' and the ``FY2022 JSPS Postdoctoral Fellowship for Research in Japan (Short-term)'', sponsored by the Japanese Society for the Promotion of Science (JSPS).
F.K.W., A.F.,
and D.E.B. acknowledge support from the joint project No. 13N15685 ``German Quantum Computer based on Superconducting Qubits (GeQCoS)'' sponsored by the German Federal Ministry of Education and Research (BMBF) under the \href{https://www.quantentechnologien.de/fileadmin/public/Redaktion/Dokumente/PDF/Publikationen/Federal-Government-Framework-Programme-Quantum-technologies-2018-bf-C1.pdf}{framework programme
``Quantum technologies -- from basic research to the market''}. D.E.B. also acknowledges support from the German Federal Ministry of Education and Research via the \href{https://www.quantentechnologien.de/fileadmin/public/Redaktion/Dokumente/PDF/Publikationen/Federal-Government-Framework-Programme-Quantum-technologies-2018-bf-C1.pdf}{framework programme
``Quantum technologies -- from basic research to the market''} under contract number 13N16210 ``SPINNING''.
F.N. is supported in part by: Nippon Telegraph and Telephone Corporation (NTT) Research, the Japan Science and Technology Agency (JST) [via the Quantum Leap Flagship Program (Q-LEAP), and the Moonshot R\&D Grant Number JPMJMS2061], the Asian Office of Aerospace Research and Development (AOARD) (via Grant No. FA2386-20-1-4069), and the Foundational Questions Institute Fund (FQXi) via Grant No. FQXi-IAF19-06.


\appendix



\section{Demonstration of coherence transfer}\label{appA}
In this section we want to analyze the origin of the oscillation frequency of both the wall and the cavity mode 2. For the sake of simplicity, we will work on the bare basis of the free Hamiltonian (see Sec.~\ref{model}), namely the eigenstates of $\hat H_0$, assuming a unitary evolution of the dynamics. Mathematical techniques employed in this appendix can be found in \cite{PhysRevA.106.033502}.

\subsection{Oscillation frequency of $X_{{\text{w}}}(t)$}
As first, we focus on the coherence transfer from mode 1 to the wall. To do this, we consider only the part of  the interaction Hamiltonian involving the cavity mode 1 and the wall. Indeed, interacting terms containing the second cavity mode play no role in the dynamics of $\hat X_{{\text{w}}}$. We consider therefore the following Hamiltonian of interaction
\begin{align}
\hat H_\textrm{I}=\frac{\hbar\omega_1\epsilon}{2}\left(\hat a_1+\hat a_1^\dag\right)^2\left(\hat b+\hat b^\dag\right).
\label{Hamiltinian:interaction:W1}
\end{align}
Now, we exploit the resonance condition $\Omega=2\omega_1$ and perform the rotating wave approximation. This choice reduces Eq.~\eqref{Hamiltinian:interaction:W1} to
\begin{align}
\hat H_\textrm{I}=\frac{\hbar\omega_1\epsilon}{2}\left((\hat{a}_1)^2\hat b^\dag+(\hat a_1^\dag)^2\hat b\right)
\label{Hamiltinian:interaction:W1:2}.
\end{align}
We remind that the cavity mode 1 is pumped in resonance $\omega_{\rm L}=\omega_1$. However, rather than adding a further Hamiltonian term for the laser, it results reasonable to assume that the laser gets the mode in a coherent state $\lvert\alpha\rangle$, with $\hat a_1\lvert\alpha\rangle=\alpha\lvert\alpha\rangle$. This choice allows us to simplify the Hamiltonian in Eq.~\eqref{Hamiltinian:interaction:W1:2} by replacing the annihilation and creation operators $\hat a_1$ and $\hat a_1^\dag$ with the coherent parameters $\alpha=F e^{-i\omega_1 t}$ and $\alpha^*=F e^{i\omega_1 t}$, respectively. At the resonance condition Eq.~\eqref{Hamiltinian:interaction:W1:2} becomes:
\begin{align}
\hat H_\textrm{I}(t)=\frac{\hbar\omega_1\epsilon\,F^2}{2}\left(e^{-i\Omega t}\hat b^\dag+e^{i\Omega t}\hat b\right).
\label{Hamiltinian:interaction:W1:3}
\end{align}
In this form, Eq.~\eqref{Hamiltinian:interaction:W1:3} describes a laser powering the mechanical mode, therefore it already witnesses the presence of coherence on the wall. We now work in the Heisenberg picture, and we write the unitary operator as $\hat U(t)=\hat U_0(t)U_\textrm{Dw}(t)$, where $\hat U_0=\exp\{-i\hat H_0 t/\hbar\}$, and 
\begin{align}
\hat U_\textrm{Dw}(t)=&\overset{\leftarrow}{\mathcal{T}}\exp\left[-\frac{i}{\hbar}\int_0^t \textrm{d}t'\hat U_0(t')\hat H_\textrm{I}\hat U_0^\dag(t')\right]=e^{\frac{i\beta t}{2}\hat X_{{\text{w}}}},
\label{DisW}
\end{align}
with $\beta=-\epsilon F^2\omega_1$ at the resonance condition.
At his point, We estimate the average value of the wall's quadrature position operator
\begin{align}
X_{{\text{w}}}(t)=&\textrm{Tr}[\hat U^\dag(t)\hat X_{{\text{w}}}\hat U(t)\rho_i]\nonumber\\
=&\textrm{Tr}[\hat U_\textrm{Dw}^\dag(t)\hat U_0^\dag(t)(\hat b^\dag+\hat b)\hat U_0(t)\hat U_\textrm{Dw}(t)\rho_i]\nonumber\\
=&\beta t\sin(\Omega t),
\label{oscwal}
\end{align}
where $\rho$ is the initial state of the system.This result shows that the wall oscillates with its own frequency, due to the presence of the coherence state of the cavity mode 1. Moreover, we want to stress the importance of taking the action of laser not detuned. If the laser was detuned, having a generic frequency $\omega_\textrm{L}$, the action of the unitary operator on the Hamiltonian interaction term $\hat U_0(t)\hat H_\textrm{I}\hat U_0^\dag(t)$ would bring a phase factor $e^{i(\Omega-2\omega_\textrm{L})t}\hat b+\textrm{h.c.}$, making $X_{{\text{w}}}(t)$ oscillate with frequency $2\omega_\textrm{L}$, but with an expected much lower intensity, proportional to $1/(2\omega_\textrm{L}-\Omega$).

This analytical approach never aimed at providing a reasonable estimation of the oscillation amplitude of $X_{{\text{w}}}(t)$. An oscillation amplitude proportional to both $t$ and $\epsilon$ cannot reflect the actual modulation of the quadrature operator, given that the presence of losses would limit its linear growing. This is evident for example from Fig.~\ref{grvseqcz}, wherein the real oscillation amplitudes of the quadratures $X_{{\text{w}}}(t)$ and $X_{{\text{2}}}(t)$, accounting for losses in both the cavity and the wall, are numerically estimated for different values of the coupling constant. Increasing $\epsilon$, the quadrature $X_{{\text{w}}}(t)$ drastically decreases, demonstrating that the linear dependence of $X_{{\text{w}}}(t)$ on $\epsilon$ actually competes with the loss mechanism.

\subsection{Oscillation frequency of $X_{{\text{2}}}(t)$}
We now want to discuss the coherence transfer from the mode 1 to the mode 2. For this purpose, we start from the Hamiltonian 
\begin{align}
\label{Hamiltinian:interaction_A6}
\hat H_\textrm{I}=&\frac{\hbar\sqrt{\omega_1\omega_2}\epsilon}{2}\left(\hat a_1+\hat a_1^\dag\right)\left(\hat a_2+\hat a_2^\dag\right)\left(\hat b+\hat b^\dag\right)\nonumber\\
&+\frac{\hbar\omega_1\epsilon}{2}\left(\hat a_1+\hat a_1^\dag\right)^2\left(\hat b+\hat b^\dag\right),
\end{align}
namely, from Eq.~\eqref{Hamiltinian:interaction} without the quadratic term for $\omega_2$. Indeed, this term would merely cause the squeezing of the second mode, without affecting  the oscillation frequency, thus, the parameter we are interested in.
As done before, we use the condition resonance $\Omega=2\omega_1$ and get rid of all fast oscillating terms by means of the rotating wave approximation. The Hamiltonian in Eq.~\eqref{Hamiltinian:interaction_A6} reduces to
\begin{align}
\hat H_\textrm{I}(t)&=\frac{\hbar\omega_1\epsilon}{2}\left(({\hat a}_1)^2\hat b^\dag+(\hat a_1^\dag)^2\hat b\right)\nonumber\\
&+\frac{\hbar\sqrt{\omega_1\omega_2}\epsilon}{2}\left(\hat a_1\hat b^\dag+\hat a_1^\dag\hat b\right)\left(\hat a_2 +\hat a_2^\dag \right).
\label{Hamiltinian:interaction:W2}
\end{align}
We adopt the same strategy proposed above: we apply a laser to the cavity mode 1 with frequency $\omega_1$, thereby substituting the relative annihilation and creation operator with $\alpha=F e^{-i\omega_1 t}$ and $\alpha^*=F e^{i\omega_1 t}$, respectively, to get
\begin{align}
\hat H_\textrm{I}(t)&=\frac{\hbar\omega_1\epsilon F^2}{2}\left(e^{-i\Omega t}\hat b^\dag+e^{i\Omega t}\hat b\right)\nonumber\\
&+\frac{\hbar\sqrt{\omega_1\omega_2}\epsilon F}{2}\left(e^{-i\omega_1 t}\hat b^\dag+e^{i\omega_1 t}\hat b\right)\left(\hat a_2+\hat a_2^\dag \right),
\label{Hamiltinian:interaction:W2:2}
\end{align}
where the first line is identical to Eq.~\eqref{Hamiltinian:interaction:W1:3}, whereas the second line describes the displacement of the second mode due to the presence of displacement in the wall. The time evolution operator has three components, $\hat U(t)=\hat U_0(t)\hat U_\textrm{Dw}(t)\hat U_\textrm{D2}(t)$: (i)  the free evolution $\hat U_0=\exp\{-i\hat H_0 t/\hbar\}$, (ii) the term $\hat U_\textrm{Dw}(t)$ corresponding to Eq.~\eqref{DisW}, (iii) and finally, 
\begin{align}
\hat U_\textrm{D2}(t)=\overset{\leftarrow}{\mathcal{T}}\exp{-\frac{i}{\hbar}\int_0^t\textrm{d}t' \hat U_\textrm{Dw}^\dag \hat U_0^\dag\hat H_\textrm{D2}(t')\hat U_0\hat U_\textrm{Dw}},
\label{UD2}
\end{align}
where $\hat H_\textrm{D2}(t)$ corrisponds to  the second line of Eq.~\eqref{Hamiltinian:interaction:W2:2}.Therefore, the unitary operator in Eq.~\eqref{UD2} is
\begin{align}
\hat U_\textrm{D2}(t)=\overset{\leftarrow}{\mathcal{T}}\exp{-\frac{i}{\hbar}\int_0^t\textrm{d}t'\hat{\tilde H}_\textrm{D2}(t')}
\label{UD2:2}
\end{align}
with
\begin{align}
\hat{\tilde H}_\textrm{D2}(t')&=\frac{\hbar\sqrt{\omega_1\omega_2}\epsilon F}{2} \left(e^{-i\omega_2 t'}\hat a_2+e^{i\omega_2 t'}\hat a_2^\dag \right)\nonumber\\
&\times\left[e^{-i\omega_1 t'}\left(\hat b-\frac{i\omega_1\epsilon F^2 t'}{2}\right)+\textrm{h.c.}\right].
\label{HD2}
\end{align}
This Hamiltonian still contains terms proportional to the wall's operators. Since we are interested in the oscillation frequency of the quadrature position operator of the second cavity mode, we can account for the time evolution of the system for large time, namely, $t\gg 2/(\omega_1\epsilon F^2)$ and reasonably ignore these terms. Hence, the Hamiltonian in Eq.~\eqref{HD2} reduces to
\begin{align}
\hat{\tilde H}_\textrm{D2}(t')=&\frac{\hbar \xi t'}{2}\sin(\omega_1 t') \left(e^{-i\omega_2 t'}\hat a_2+e^{i\omega_2 t'}\hat a_2^\dag \right)
\label{tildehd2}
\end{align}
where we introduced $\xi=-\omega_1 \sqrt{\omega_1\omega_2}\epsilon^2 F^3$. 

The time evolution of the quadrature operator of the second mode can be estimated in Heisenberg picture:
\begin{align}
X_{\textrm{2}}(t)=&\textrm{Tr}[\hat X_{\textrm{2}}^{\textrm{H}}(t)\rho_i],
\label{x2}
\end{align}
with $\hat X_{\textrm{2}}^{\textrm{H}}(t)=\hat U^\dag(t)\hat X_{\textrm{2}}\hat U(t)$, and $\hat X_{\textrm{2}}=\hat a_2+\hat a_2^\dag$.
For our purpose, we estimate the unitary operator in Eq.~\eqref{UD2:2} as 
\begin{align}
\hat U_\textrm{D2}(t)=\exp\left[\frac{\xi}{4}\left[K(\omega_2-\omega_1,t)-K(\omega_2+\omega_1,t)\right]\hat a_2^\dag-\textrm{h.c.}\right]
\end{align}
modulo an overall complex phase that has no physical significance, and 
\begin{align}
K(\omega,t)=&\int_0^t\textrm{d}t't'e^{-i\omega t'}=\frac{e^{-i\omega t}(1+i\omega t)-1}{\omega^2}
\end{align}
Finally, one obtains
\begin{align}
\hat X_{\textrm{2}}^{\textrm{H}}(t)&=\hat U_\textrm{D2}^\dag(t)\hat U_\textrm{Dw}^\dag(t)\hat U_0(t)\hat X_{\textrm{2}}^\dag\hat U_0(t)\hat U_\textrm{Dw}(t)\hat U_\textrm{D2}(t)\nonumber\\
&= e^{-i\omega_2 t}\frac{it \xi}{4}\left[\hat a_2+\frac{\,e^{i(\omega_2-\omega_1)t}}{\omega_2-\omega_1} -\frac{ e^{i(\omega_2+\omega_1)t}}{\omega_2+\omega_1}\right.\nonumber\\
&+e^{i(\omega_2-\omega_1)t/2}\frac{\sinc\left[(\omega_2-\omega_1)t/2\right]}{\omega_2-\omega_1}\nonumber\\
&\left.-e^{i(\omega_2+\omega_1)t/2}\frac{\sinc\left[(\omega_2+\omega_1)t/2\right]}{\omega_2+\omega_1}\right]+\textrm{h.c.}\; ,
\end{align}
and, from Eq.~\eqref{x2},
\begin{align}
X_{\textrm{2}}(t)&=\frac{\xi t}{2}\left\{\left(\frac{1}{\omega_2+\omega_1}-\frac{1}{\omega_2-\omega_1}\right)\sin(\omega_1 t)\right.\nonumber\\
&+\frac{\sin\left[(\omega_2+\omega_1)t/2\right]\sinc\left[(\omega_2-\omega_1)t/2\right]}{\omega_2-\omega_1}\nonumber\\
&-\left.\frac{\sin\left[(\omega_2-\omega_1)t/2\right]\sinc\left[(\omega_2+\omega_1)t/2\right]}{\omega_2+\omega_1}\right\}\; .
\label{osc2}
\end{align}

Notice that, the last line contains the function $\sinc(x)=\sin(x)/x$, therefore, its contribution is negligible as soon as $t\gg 1/(\omega_2-\omega_1)$.
Note that the oscillation frequency of $X_{\textrm{2}}(t)$ does not depend on $\omega_2$. This means that every cavity mode oscillates with frequency $\omega_1$ and amplitude proportional to $\sqrt{\omega_n}/(\omega_n-\omega_1)$.

\subsection{Dependence of $X_{\textrm{2}}(t)$ on $\epsilon$}
So far we were interested in the oscillation frequency of the position quadrature operator of mode 2. For this scope, the unitary evolution employed so far was sufficient. This is due to the fact that, at least in the regime this analysis is based on, the presence of losses does not affect the oscillation frequency of the quadrature operator but only its amplitude. Nevertheless, we can still ask what is the expected dependence of $X_{\textrm{2}}(t)$ on $\epsilon$, in order to make a comparison with $X_{\textrm{w}}(t)$. According to our findings in Fig.~\ref{lcvshccz}, this dependence must generally play a stronger role. This was already seen in Eq.~\eqref{osc2}, where the oscillation amplitude increases quadratically with respect to $\epsilon$. 

To have a further hint, we now account for the whole Hamiltonian in Eq.~\eqref{Hamiltinian:interaction}, namely we includes the up- and down-conversion terms between cavity mode 2 and the wall. Following the same procedure employed so far, after some  calculations we find that up- and down-conversion terms contributes to the time evolution of $X_{\textrm{2}}(t)$ with a further net displacement:
\begin{align}
\label{U}
\hat U_{\textrm{2w}}=&\overset{\leftarrow}{\mathcal{T}}\exp{-\frac{i}{\hbar}\int_0^t\textrm{d}t' \hat U_\textrm{Dw}^\dag \hat U_\textrm{D2}^\dag\hat U_0^\dag\hat H_\textrm{2w}(t')\hat U_0\hat U_\textrm{D2}\hat U_\textrm{Dw}}\nonumber\\
\simeq&\exp\left\{\frac{\xi^2t^2}{8}\hat a_2^\dag\left[\frac{\,e^{i(\omega_2-\omega_1)t}}{\omega_2-\omega_1} -\frac{ e^{i(\omega_2+\omega_1)t}}{\omega_2+\omega_1}\right]-\textrm{h.c.}\right\}\;,
\end{align}
where
\begin{align}
\hat H_\textrm{2w}(t)=\hbar\epsilon\frac{\omega_2}{2}\left(\hat a_2+\hat a_2^\dag\right)^2\left(\hat b+\hat b^\dag\right)\;.
\end{align}

Letting the unitary operator in Eq.~\eqref{U} act on Eq.~\eqref{osc2}, we obtain
\begin{align}
X_{\textrm{2}}(t)=&\langle\hat U_{\textrm{2w}}^\dag\hat U_\textrm{D2}^\dag\hat U_\textrm{Dw}^\dag\hat U_0^\dag\hat X_{\textrm{2}}\hat U_0\hat U_\textrm{Dw}\hat U_\textrm{D2}\hat U_{\textrm{2w}}\rangle\nonumber\\
\simeq& \frac{\xi^3 t^3}{16}\left(\frac{1}{\omega_2+\omega_1}-\frac{1}{\omega_2-\omega_1}\right)\sin(\omega_1 t)+O(\xi t).
\label{osc2new}
\end{align}
Although the scope of this analysis is not to accomplish a real estimation of the oscillation amplitude of $X_{\textrm{2}}(t)$, it roughly provides a benchmark for the comparison with $X_{\textrm{w}}(t)$, showing a strong dependence of $X_{\textrm{2}}(t)$ on $\epsilon$.

\bibliography{alessandria}

\end{document}